# An ethnographic study of visual impairments for voice user interface design

**Gisela Reyes-Cruz**
University of Nottingham
Nottingham, UK
psxagre@nottingham.ac.uk

**Joel Fischer**
University of Nottingham
Nottingham, UK
joel.fischer@nottingham.ac.uk

**Stuart Reeves**
University of Nottingham
Nottingham, UK
stuart.reeves@nottingham.ac.uk

## ABSTRACT

Design for Voice User Interfaces (VUIs) has become more relevant in recent years due to the enormous advances of speech technologies and their growing presence in our everyday lives. Although modern VUIs still present interaction issues, reports indicate they are being adopted by people with different disabilities and having a positive impact. For the first author's PhD research project, an ethnographic study is currently being carried out in a local charity that provides support and services to people with visual impairments. The purpose is to understand people's competencies and practices, and how these are, or could be, related to voice technologies (assistive technology and mainstream VUIs). Through direct observation and contextual interviews, we aim to

investigate the problems and solutions they encounter and the ways they cope with particular situations.

## INTRODUCTION

Over the past years, Voice User Interfaces (VUIs) have become widely available to the general public in mobile devices, home assistants, developed for controlling 'smart homes', embedded in wearables and recently distributed for in-car use. Despite these substantial efforts to make VUIs ubiquitous in our daily lives, there are still some interaction issues that need to be addressed. Primarily, these VUIs have been designed for one-off interactions, not really providing a conversational experience. As real-world settings tend to be complex, with many people speaking and many activities happening at the same time, user expectations when using VUIs are not met [6,7]. Furthermore, reports indicate that VUIs are commonly used for the same simple tasks (e.g. music, short queries, clock functions), suggesting poor usability and poor discoverability [3].

Past work on speech technology has lacked a strong involvement of Human Computer Interaction (HCI) research. It was not until recently that work has been done around discovering and understanding the practical actions and reasoning of people using VUIs in social settings such as the home [7]. Likewise, the role of VUIs in accessibility has not been investigated in-depth outside the assistive technology field. Research has noted that use of assistive technology sometimes evokes frustration and self-consciousness [8,15], thus the call for improving accessibility of mainstream technologies. Work has been done aiming to provide digital accessibility to people with different disabilities: design guidelines have been outlined and efforts to raise accessibility awareness have been increasing, but most of this work has been tailored to improve online websites. Moreover, people with disabilities still find several barriers when accessing digital content [16]. Additionally, with the rise of mobile and ubiquitous computing, work has been aiming to expand on the notion of context in designing systems by considering its three dimensions: characteristics of humans, environments and applications [11]. Research acknowledges that there are very close parallels between problems faced by people with physical or cognitive disabilities in a typical situation and those faced by people without permanent disabilities in constraining or encumbered circumstances (e.g. walking in public while texting), generally called situationally-induced impairments and disabilities (SIIDs) [11, 17]. Previous work on this space outlined factors causing SIIDs during interaction with mobile devices, such as ambient characteristics (e.g. temperature, noise) and user's state (e.g. in motion, mood), but there are some research gaps that need to be addressed to better understand the origins and effects of SIIDs [9].

The first author's PhD research aims to broaden empirical work around VUIs in everyday life drawing on ethnography in order to understand first-hand the needs of, and challenges presented to, people with disabilities in their day-to-day lives, as well as learning about their use of voice user interfaces. Along these lines, future research will aim to explore possible solutions that take into consideration findings from this fieldwork and contrasting them with applications that could be useful to wider audiences.

**DESIGNING FOR ACCESSIBILITY**
The latest world report on disability [18] by the World Health Organization estimates that a 15% of the entire world population has some form of disability. Additionally, elder population in the world is expected to increase dramatically in the next decades [19]. Thus, different design approaches and frameworks have been outlined in order to improve physical and digital accessibility (e.g. Universal Design, Inclusive Design, Ability-based Design) [17], Some research has also explored the similarities and differences between permanent, temporary and situational disabilities, as well as outlined the benefits of such intersectional design for improving general accessibility and usability of visual interfaces. For example, prior work on SIIDs has reported similar performances of people using small screens or mobiles while walking and people with motor impairments using desktop or mobile devices [9,14]. Experimental studies have also indicated that task performance decreases due to visual (e.g. low screen brightness) and cognitive (e.g. music in background) situational impairments [5,10].

Nevertheless, people with a diverse range of disabilities are not normally considered for interface design and research, and situational disabilities they experience on top of their current impairments (i.e. due to specific environments or contexts) have not been considered in-depth [1, 9]. Furthermore, some situational impairments such as those caused by ambient noise or user mood are still underexplored in research. Thus, the need for a better and deeper understanding of the nature and effects of different situational impairments [5,9,10,14].

**ACCESSIBLE VOICE INTERFACES**
Mainstream voice assistants, such as *Alexa* (Amazon Echo) and *Siri* (Apple), have been adopted by people with different disabilities, having an especially positive impact on the elderly, and the visual, motor and cognitive impaired [2,4,8]. Despite VUIs current limitations, in many cases they provide these users new ways to gain more independence and control over their daily activities, and they are reported as easy to use and easy to learn [19]. In some cases, VUIs have been appropriated by people with disabilities and repurposed in ways not intended by designers (e.g. using *Alexa* for blinking lights in other rooms to get someone's attention [8]). Although modern VUIs available to the general public represent an opportunity to improve the lives of people with different impairments, interaction support and accessibility were not a primary interest of companies behind them, as opposed to improving speech recognition APIs by large scale data collection.

On the other hand; in the assistive technology field, speech-based interfaces have been widely implemented for supporting people with disabilities and helping them accomplish digital tasks that they would find hard or impossible to perform otherwise. The two most common implementations are screen readers (e.g. *JAWS* for desktop, *VoiceOver* for mobile) and speech recognition tools (e.g. *Dragon NaturallySpeaking*) [8,15]. However, some issues with assistive technologies arise as most of them are an intermediate layer between the original interface and the user. Screen readers have become a fundamental tool for blind and visual impaired people, but their successful use relies not on the tool itself but on the structure of the underlying website or system, as many of these do not meet accessibility guidelines resulting in complicated and cumbersome interactions. Furthermore, interaction is even more difficult for people with combined disabilities, as for instance motor impairments do not always allow them to perform specific gestures required by some screen readers [4,13].

We build on some general principles of inclusive design approaches by a) including people with different impairments into this research and design process, b) aiming to understand their competencies and practices, and c) focusing on specific enduring impairments in order to apply findings to temporary or situational impairments to the greatest extent possible.

**UNDERSTANDING THE LIVES OF PEOPLE WITH VISUAL IMPAIRMENTS THROUGH ETHNOGRAPHY**
Ethnography's main emphasis is direct observation, *in situ*. The researcher is immersed in the setting, participating with members of the group being studied and having the opportunity to see what they do at first hand on their own terms. Within flexible design studies, ethnography is the most appropriate way for obtaining an in-depth understanding of a group of people of interest, or about specific aspects of their lives, by focusing on their behaviour, interactions, language and artefacts [12]. Moreover, ethnomethodological and conversation analytic (EMCA) approaches have proved useful for uncovering the interactional methods employed by people when talking to voice assistants [7]. EMCA-informed research concentrates on articulating the organisation of social action, moment by moment. This approach could be valuable to ascertain particular skills and methods used by people with different impairments in their daily activities, some of them including the use of assistive and/or voice technologies.

Fieldwork of this study is being conducted at the local branch of a charity that supports blind and partially sighted people in the UK by providing them with a range of services (e.g. social meetings, counselling, IT guidance). The initial weeks have consisted of researcher immersion into the charity community by approaching staff members acting as gatekeepers in order to have access to potential participants. As a result of this contact the main researcher attended a Sight Loss Awareness and Sight Guide training session offered by the charity, which provided introductory information about different eye conditions; recommendations to communicate effectively with people living with sight

*Some people in the meetings use lens magnifiers, carried in their pockets, to read printed text or text on their phones. Some use oversized fonts on computers or mobile phones, while others have to move the device close to their face and tilt their head in precise angles in order to read the text. A few people expressed their use of voice assistants (Siri, Alexa and Google) as a personal preference, while for others such assistants and screen readers are the only alternative to interact with technology due to the severity of their visual impairment. More specifically, one person said he preferred Android screen readers over Apple because the gestures are simpler, whereas a blind participant prefers a wireless keyboard connected to the phone for using VoiceOver (screen reader for iPhone).*
**Vignette 1. Diversity in the study group**

a) Identify yourself – don't assume the person will recognise you by your voice.
b) Name the person when introducing yourself or when directing conversation to them in a group.
c) When leaving a room or a conversation, let the person know it.

**Vignette 2. General recommendations to communicate with people who are visually impaired**

loss; the impact this has on their lives; barriers to independence; and emotional consequences. Furthermore, 'simulation spectacles' were used to illustrate different visual limitations and paired work using blindfolds was employed to teach sighted guiding in different contexts (e.g. indoors, outdoors) and negotiation in different situations (e.g. stairs, doors).

Follow-up actions have consisted of attending social and IT meetings held at the charity offices as part of being immersed into the community and building rapport. The charity is composed of staff members, volunteers and service users. These groups are integrated by both sighted people and people with different types of visual impairments. Many service users are older adults (over 60 years old) and staff and volunteers are from a younger demographic (20 to 60 years old). Social meetings at the charity are informal gatherings held every fortnight where volunteers and service users get together to talk and have a drink. IT meetings are drop-in sessions that happen twice a week to provide guidance to people with visual impairments on how to use technology, including smartphones, computers and different types of assistive technology; however, these sessions also serve as community gatherings for regular attendees, thus the teaching is embedded within the social context of the group. Informal conversations and observation of some activities (e.g. teaching and use of screen readers) have taken place in these meetings and data has been collected through fieldnotes.

**EARLY INSIGHTS FROM ETHNOGRAPHIC APPROACH**
Building on our early observations from this ethnographic study we summarize key insights that could be explored in future research about people with visual impairments, and how to design VUIs that support their activities and contexts.

*Disabilities are different for every individual.*
Attending the charity meetings has allowed us to meet people with diverse range of visual impairments, some more severe than others. For instance, conditions such as cataracts cause blurred or misty vision, while others such as glaucoma cause loss of peripheral vision. Therefore, service users have different needs, preferences and methods. Vignette 1 describes our observations on this matter, illustrating the diversity of abilities and predilections of the study group. Moreover, some people from the community possess combined impairments: in addition to the vision loss, they can exhibit different types of motor and hearing impairments; thus, preferences are further shaped by these individual characteristics.

*Interacting with visually impaired people.*
Methods of communication and engagement with people in the charity also vary depending on their eye condition. For instance, participants who are blind or almost blind require conversational vocal cues (e.g. utterances) that replace visual ones (e.g. nodding) to indicate they have been heard

*1. A blind person using JAWS (screen reader for desktop) accessed the mobile versions of websites like Facebook and YouTube in the desktop browser because they are less cluttered and therefore easier to use.*
*2. A partially sighted person said he goes to the shop when it opens (around 6 am) because it is normally empty, so to avoid other customers.*
*3. Many people have acquired accessible watches through the charity. These speak the time and date aloud when a button is pressed.*
*4. The charity records local news on CDs and sends them through the post to service users that request them.*

**Vignette 3. Examples of methods and tools used by the study group**

and understood, so they can continue talking; on the other hand, communication with people who are partially sighted still relies on body language. Vignette 2 describes some considerations to communicate effectively with people who are blind or vision impaired. Learning from these interactions could help us explore different alternatives to implement VUI solutions.

*Methods and workarounds used by people with visual impairments in their daily lives.*
People from the charity employ different practices for getting through physical and digital spaces, as well for staying connected with the social environment. Likewise, they make use of different tools to accomplish their goals. Vignette 3 describes some examples of these methods and workarounds including speech and audio applications.

*Attitudes and expertise of the study group.*
It became apparent during our fieldwork that, for the IT sessions, neither purpose of attendance nor level of proficiency was uniform. Learning to use technology, practicing use of it, leisure activities (e.g. browsing the web) and socializing with the group were all reasons given for attendance. Moreover, most of the group are novice or intermediate while staff members are more proficient, as they are in charge of teaching - as mentioned before, some staff and volunteers are also visually impaired, which allows them to give specific guidance and recommendations based on personal experience. It is important to note that although many service users are older adults, we have observed different attitudes and technological competence within the age group. People who feel more confident using technology have expressed their preferences for audible podcasts and YouTube as learning resources.

*Social and emotional implications related to technology adoption.*
People attending social meetings do not normally attend IT sessions; one person expressed her own personal point of view on this as follows. Similar to the experience of other service users, sight loss has been a gradual process for her: a number of years separate her initial symptoms and diagnosis from complete blindness. While she needs to adapt to the emotional and social day-to-day factors arising from the visual impairment, she stated that learning to use technology on top of everything else would be overwhelming, despite hearing positive comments about the Amazon Echo from another service user. Her general belief is that technological literacy before sight loss plays an important role in assistive technology adoption.

On the other hand, a regular IT session attendee voiced positive comments about *Siri*. As his visual impairment is not too severe, he uses oversized fonts for computer and mobile instead of a screen reader. However, he finds the voice assistant particularly convenient for daily tasks he finds difficult to perform otherwise (e.g. scheduling events, setting reminders).

Notably, service users and volunteers indicated the important role the charity plays in their lives. Besides facilitating the development of social and emotional support between people with similar eye conditions, they value the opportunity to explore what type of technology is more suitable for them and how to use it. They have also noted a positive impact on independence and isolation. Nevertheless, a couple of respondents mentioned a financial barrier to obtaining assistive technology as some items are considered expensive (e.g. desktop screen readers, accessible phones).

**CONCLUSIONS AND FUTURE WORK**

In this paper we described an ongoing ethnographic study as part of the first author's PhD research. The purpose is uncovering methods used and activities carried out by people with visual impairments in their daily lives. We described insights from early observations at a local branch of a charity that supports blind and partially sighted people in the UK, highlighting the complexity of their lives. Future research should address diversity of abilities, different levels of expertise, and demographics. Planned future work will include one-on-one interviews with members of the charity and observational studies of specific activities, involving shadowing participants for a part or whole of a day. With an ethnomethodological and conversational analytic approach we intend to identify possible areas to implement or improve voice user interface applications, in terms of accessibility and overall user experience. In addition, there is an opportunity for contribution to the field of SIIDs by exploring similarities to experiences of people who are blind or partially sighted when using mobile voice technologies and how factors such as ambient noise or social context affect interaction.